\documentclass[a4paper,fleqn]{cas-sc}

\usepackage[numbers, sort&compress, square]{natbib}

\usepackage{lineno,hyperref}
\usepackage{amsmath}
\allowdisplaybreaks[1]
\usepackage{graphicx}
\usepackage{float}
\usepackage{geometry}
\usepackage{subfig}
\usepackage{longtable}
\usepackage{xcolor}
\usepackage{soul}
\sethlcolor{yellow}
\modulolinenumbers[5]

\def\tsc#1{\csdef{#1}{\textsc{\lowercase{#1}}\xspace}}
\tsc{WGM}
\tsc{QE}
\tsc{EP}
\tsc{PMS}
\tsc{BEC}
\tsc{DE}

\begin{document}
\let\WriteBookmarks\relax
\let\printorcid\relax

\shorttitle{}

\shortauthors{Yishen Jiang et al.}

\title [mode = title]{Nonlinear Public Goods Game in Dynamical Environments}                      


\author[1,4,6]{Yishen Jiang}
\author[2,4,6,7,9]{Xin Wang}
\author[2,4,6]{Wenqiang Zhu}
\author[1,4,6]{Ming Wei}
\author[2,4,6,7]{Longzhao Liu}\ead{longzhao@buaa.edu.cn}\cormark[1]
\author[2,3,4,5,6,7,8]{Shaoting Tang}
\author[10]{Hongwei Zheng}\ead{hwzheng@pku.edu.cn}\cormark[2]

\affiliation[1]{organization={School of Mathematical Sciences, Beihang University},
            city={Beijing},
            postcode={100191},
            country={China}}

\affiliation[2]{organization={School of Artificial Intelligence, Beihang University},
            city={Beijing},
            postcode={100191},
            country={China}}

\affiliation[3]{organization={Hangzhou International Innovation Institute, Beihang University},
            city={Hangzhou},
            postcode={311115},
            country={China}}

\affiliation[4]{organization={Key Laboratory of Mathematics, Informatics and Behavioral Semantics, Beihang University},
            city={Beijing},
            postcode={100191},
            country={China}}

\affiliation[5]{organization={Institute of Medical Artificial Intelligence, Binzhou Medical University},
            city={Yantai},
            postcode={264003},
            country={China}}

\affiliation[6]{organization={Zhongguancun Laboratory},
            city={Beijing},
            postcode={100094},
            country={China}}

\affiliation[7]{organization={Beijing Advanced Innovation Center for Future Blockchain and Privacy Computing, Beihang University},
            city={Beijing},
            postcode={100191},
            country={China}}

\affiliation[8]{organization={Institute of Trustworthy Artificial Intelligence, Zhejiang Normal University},
            city={Hangzhou},
            postcode={310013},
            country={China}}

\affiliation[9]{organization={State Key Laboratory of General Artificial Intelligence, BIGAI},
            city={Beijing},
            postcode={100080},
            country={China}}

\affiliation[10]{organization={Beijing Academy of Blockchain and Edge Computing},
            city={Beijing},
            postcode={100085},
            country={China}}



\cortext[cor1]{First corresponding author.}
\cortext[cor2]{Second corresponding author.}

\begin{abstract}
The evolutionary mechanisms of cooperative behavior represent a fundamental topic in complex systems and evolutionary dynamics. Real-world collective interactions, particularly in multi-agent systems, are often characterized by behavior-dependent mechanism switching where the environmental state is endogenously shaped by group strategies. However, existing models typically treat such environmental variations as static stochasticity and neglect the closed-loop feedback between environmental states and cooperative behaviors. Here, we introduce a dynamic environmental feedback mechanism into a nonlinear public goods game framework to establish a coevolutionary model that couples environmental states and individual cooperative strategies. Our results demonstrate that the interplay among environmental feedback, nonlinear effects, and environmental randomness can drive the system toward a wide variety of steady-state structures, including full defection, full cooperation, stable coexistence, and periodic limit cycles. Further analysis reveals that asymmetric nonlinear parameters and environmental feedback rates exert significant regulatory effects on cooperation levels and system dynamics. This study not only enriches the theoretical framework of evolutionary game theory but also provides a foundation for modeling environmental feedback loops in scenarios ranging from ecological management to the design of cooperative mechanisms in autonomous systems.
\end{abstract}


\maketitle

\section{Introduction}
\label{introduction}

Unraveling the mechanisms that sustain cooperation in the long-term evolution of complex systems remains a central open question in recent quantitative biology and social systems research~\cite{leimar2023game, van2022human}. The dilemma stems not merely from free-riding behaviors induced by the misalignment between individual and collective interests but also from the ubiquitous uncertainty and multi-agent higher-order interactions inherent in collective actions which can alter payoff structures and interaction rules thereby significantly reshaping population dynamics~\cite{gao2024social, battiston2021physics}. To characterize such multi-agent dilemmas evolutionary game theory provides a unified dynamical framework for describing the conditions under which cooperation emerges and persists~\cite{traulsen2023future, glynatsi2024conditional, michel2024evolution}. While foundational insights were largely derived from pairwise interactions contemporary research increasingly employs the Public Goods Game as a canonical baseline for multiplayer collective action~\cite{perc2017statistical, wei2025indirect, zhu2025evolution}. In this framework individual contributions to a public pool are aggregated and synergistically enhanced before being redistributed among all participants.

Despite its mathematical simplicity, the linear PGG cannot capture key nonlinear features of real group interactions in many biological and social systems. Examples include the production of extracellular enzymes in microbial populations, where the metabolic benefit may accelerate with more contributors, or saturate due to limited uptake or spatial constraints~\cite{fritts2021extracellular}. Similarly, in human societies, the productivity of a collective endeavor may exhibit increasing returns with scale or saturate and even decline due to capacity limitations and congestion~\cite{hauert2024frequency}. A foundational step toward incorporating such realism was the introduction of nonlinear public goods games, which modeled synergy and discounting via a nonlinear benefit function~\cite{hauert2006synergy, zhu2024reputation}. Subsequent work developed general tools for analyzing arbitrary benefit shapes and clarified how nonlinearity reshapes equilibria and dynamics in multiplayer cooperation~\cite{archetti2018analyze}.

While nonlinear models better reflect the graded nature of real-world interactions, most remain deterministic and miss the environmental variability in real systems~\cite{cressman2014replicator}. Crucially, various forms of stochasticity, including behavioral noise, random participation, and extrinsic environmental shifts, can be unified within the framework of discrete environmental changes~\cite{unakafov2020emergence}. In such models, the system switches between distinct states , each characterized by specific payoff structures or interaction rules~\cite{hilbe2018evolution, su2019evolutionary, taitelbaum2023evolutionary}. This switching profoundly alters evolutionary dynamics, affecting the stability of cooperative equilibria and enabling transitions unattainable in static settings~\cite{feng2022stochastic, luo2025evolutionary}. The interaction between nonlinear social dilemmas and such discrete environmental variability remains a frontier. Recent work shows that when a nonlinear PGG stochastically alternates between synergistic and discounting regimes, cooperation can be sustained in parameter ranges that would not support it in static settings~\cite{zhu2024evolutionary}. These observations motivate integrating nonlinearity with environmental randomness to explain cooperation in realistic contexts.

However, models based on discrete environmental switching largely overlook the continuous nature of environmental change in the real world. In natural and social systems, variables such as resource abundance, risk levels, institutional strength, and social trust often vary continuously, driving evolutionary outcomes away from those predicted by static averaging~\cite{wang2025evolutionary, salles2024evolution}. Periodic fluctuations or stochastic perturbations in the external environment can also induce substantial oscillations in cooperation levels, prompting switches between states and producing nonstationary dynamics~\cite{taitelbaum2023evolutionary, gokhale2016eco}. Studies in ecology and sociology have further shown that environmental states and group behaviors interact through bidirectional feedback, where resource abundance shapes individual strategic choices and individual behaviors, in turn, alter environmental conditions~\cite{weitz2016oscillating, tilman2020evolutionary, hua2024coevolutionary, ma2024effect}. Specifically, environmental feedback can sustain and promote cooperation by modulating time scales~\cite{shao2019evolutionary}, enabling manifold control~\cite{wang2020steering}, and reshaping network structure~\cite{betz2024evolutionary}. Furthermore, the coupling between local strategy-dependent feedback and global environmental fluctuations can shape long-term dynamics under different coupling patterns, expanding the parameter regions that support cooperation across multiple time scales~\cite{jiang2023nonlinear}.

Public goods games in the real world often take place in environments that are themselves constantly changing~\cite{barrett2013climate, zubrickas2014provision, chen2019imperfect}. From nutrient abundance influenced by metabolites in microbial populations~\cite{kramer2020bacterial} to social trust relationships that fluctuate with cooperative behaviors in human societies~\cite{kumar2020evolution}, the environmental state continuously coevolves with collective strategies. This feature is particularly critical in modern multi-agent collaborative systems~\cite{cimpeanu2022artificial}. Take the Internet of Vehicles or unmanned system swarms as a prime example where individual agents must trade off between sharing critical information and free-riding while the system-level communication protocol dynamically switches between a synergy mode and a conservative defense mode based on monitoring of collective cooperation levels ~\cite{alsaqabi2023incentivizing, alalwany2024security, wu2022game}. This behavior-dependent mechanism switching renders the environmental state not as an exogenously given static background or discrete noise but creates a complex feedback control problem, yet existing theoretical models have failed to effectively capture this continuous and bidirectional dynamic regulation mechanism.

To quantitatively analyze such mechanism switching processes driven by collective behaviors and address the aforementioned feedback control challenge, we construct a nonlinear public goods game model incorporating continuous dynamic feedback. We endogenize the environmental state as a dynamic variable regulated by the abundance of cooperators, thereby establishing coupled equations between strategy evolution and the environment. This framework captures not only the unidirectional selection pressure of the environment on strategies but also the reverse reshaping effect of strategies on the environment.

Generally, the main novel results and potential applications of this study are summarized as follows:

\begin{itemize}
    \item We construct a eco-evolutionary framework that couples nonlinear public goods games with continuous environmental feedback. Unlike models with static stochasticity, this approach captures the closed-loop interaction where collective strategies endogenously reshape the environment.
    
    \item We theoretically identify that the mismatch between feedback sensitivity and evolutionary speed is sufficient to induce limit cycle oscillations, while asymmetric payoff parameters can effectively reshape the boundaries of cooperation to stabilize the system.
    
    \item The proposed framework offers broad applicability to real-world scenarios, including protocol switching in multi-agent systems, the management of recurrent waves in large-scale health crises, and the regulation of charitable resource flows.
\end{itemize}

The structure of this paper is organized as follows: The subsequent sections formulate the mathematical model incorporating dynamic feedback in Sec.~\ref{sec2}. Subsequently, Sec.~\ref{sec3} investigates the evolutionary dynamics under static, symmetric, and asymmetric settings using combined theoretical and numerical approaches. Finally, Sec.~\ref{sec4} summarizes the main findings and discusses the practical implications of the model.

\section{Model}\label{sec2}
In the traditional PGG model, each round of the game involves randomly selecting $N$ participants from a well-mixed, infinitely large population. Each participant i can choose between two strategies: cooperation (C) and defection (D). Each cooperator contributes an amount $c$ to the common pool, while defectors contribute nothing. For simplicity without loss of generality, we set $c = 1$ throughout the paper. Suppose that there are $n_\mathrm{C}$ cooperators in the group, so the total contribution to the common pool is $n_\mathrm{C}$. This total contribution is multiplied by a multiplication factor $r$ $(r>0)$ and evenly distributed among all participants in the group. In other words, the traditional PGG describes a linear relationship between the payoff and the number of cooperators.

In the nonlinear PGG model, the actual payoff depends nonlinearly on the number of cooperators and the total contribution to the common pool, which has been documented in both biological and economic contexts~\cite{archetti2012game}. Here, we introduce a nonlinear parameter $\omega$ to capture the effects of synergy and discounting~\cite{hauert2006synergy}. When there are $n_\mathrm{C}$ cooperators ($n_\mathrm{C} > 0$) in a group of size $N$, the payoffs for cooperators and defectors can be expressed as
\begin{align}\label{eq_Pi_CD}
\begin{split}
	\pi_\mathrm{D}(n_{\mathrm{C}}) &=\frac {r}{N}(1+\omega+\omega^2+\cdots+\omega^{n_{\mathrm{C}}-1})=\frac {r}{N}\frac{1-\omega^{n_{\mathrm{C}}}}{1-\omega}, \\
	\pi_\mathrm{C}(n_{\mathrm{C}}) &=\pi_\mathrm{D}(n_{\mathrm{C}})-1, 
\end{split}
\end{align}
where $\omega$ determines whether each additional cooperator produces a higher return ($\omega > 1$, synergy) or a lower return ($\omega < 1$, discounting). When $\omega = 1$, Eq.  (\ref{eq_Pi_CD}) reduces to the traditional linear PGG case.


Most previous studies have assumed that the common pool exhibits either synergy or discounting, but not both. However, a recent study has explored scenarios in which both types of nonlinear effects coexist~\cite{zhu2024evolutionary}. Inspired by this approach, we assume that with probability $p$ $(0<p<1)$, the group participates in a discounting PGG (dPGG) and with probability $1 - p$, it engages in a synergistic PGG (sPGG). To further distinguish between the two cases, we introduce a parameter $\delta$ ($0 < \delta < 1$). Specifically, we set $\omega = 1 + \delta$ for sPGG and $\omega = 1 - \delta$ for dPGG. Furthermore, asymmetry in interactions is widespread in real-world systems\cite{li2025asymmetric}. To account for such asymmetry, we now introduce two separate parameters, $\delta_d$ and $\delta_s$, to more precisely distinguish the nonlinear effects in dPGG and sPGG, respectively. Accordingly, the payoffs for cooperators and defectors under the sPGG and dPGG settings can be expressed as
\begin{equation}\label{eq_Pi_sd_asy}
\begin{aligned}
\pi_{\mathrm{C}}^{\mathrm{sPGG}}(n_{\mathrm{C}}) &= \frac{r}{N} \cdot \frac{(1+\delta_s)^{n_\mathrm{C}} - 1}{\delta_s} - 1, \\
\pi_{\mathrm{D}}^{\mathrm{sPGG}}(n_{\mathrm{C}}) &= \frac{r}{N} \cdot \frac{(1+\delta_s)^{n_{C}} - 1}{\delta_s}, \\
\pi_{\mathrm{C}}^{\mathrm{dPGG}}(n_{\mathrm{C}}) &= \frac{r}{N} \cdot \frac{1 - (1-\delta_d)^{n_{\mathrm{C}}}}{\delta_d} - 1, \\
\pi_{\mathrm{D}}^{\mathrm{dPGG}}(n_{\mathrm{C}}) &= \frac{r}{N} \cdot \frac{1 - (1-\delta_d)^{n_{\mathrm{C}}}}{\delta_d}.
\end{aligned}
\end{equation}

Moreover, the parameter $p$ operates at the group level. In addition, we assume that the property of the common pool, whether it follows sPGG or dPGG, is randomly determined when calculating payoffs. Therefore, given $n_\mathrm{C}$ cooperators in the group, the expected payoffs for cooperators and defectors can be expressed as

\begin{align}\label{eq_Pi_ave}
	\begin{split}
		\pi_{\mathrm{C}}(n_{\mathrm{C}},p)&=(1-p)\pi_{\mathrm{C}}^{\mathrm{sPGG}}(n_{\mathrm{C}})+p\pi_{\mathrm{C}}^{\mathrm{dPGG}}(n_{\mathrm{C}}),\\
		\pi_{\mathrm{D}}(n_{\mathrm{C}},p)&=(1-p)\pi_{\mathrm{D}}^{\mathrm{sPGG}}(n_{\mathrm{C}})+p\pi_{\mathrm{D}}^{\mathrm{dPGG}}(n_{\mathrm{C}}).
    \end{split}
\end{align}

\begin{figure}
    \centering
    \includegraphics[width=0.75\textwidth]{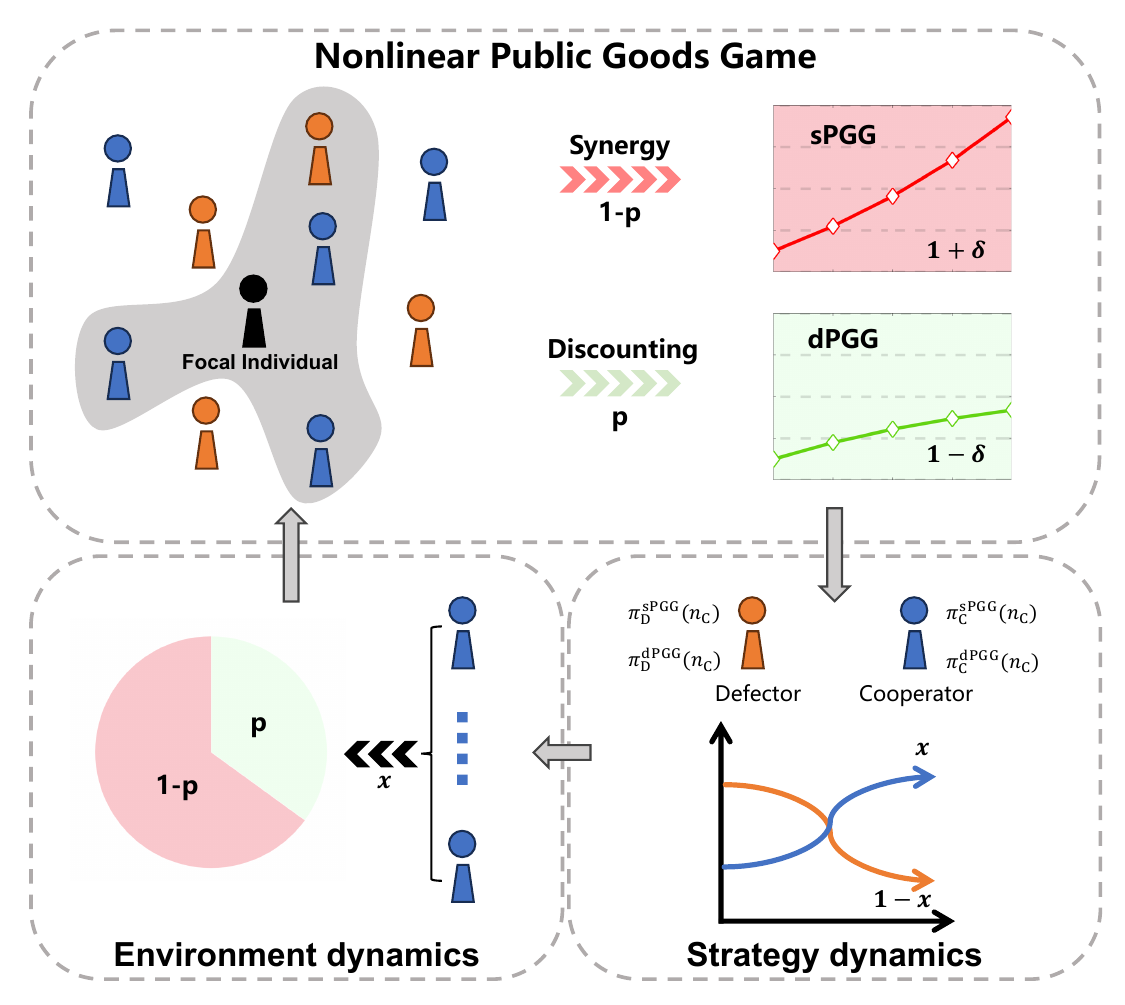}
	\caption{\textbf{Model schematic}. In a well-mixed population, groups of size $N$ are formed randomly each round to play a nonlinear PGG, with strategy evolution governed by replicator dynamics. The environmental state $p$ coevolves with population composition, modulating game outcomes probabilistically: each interaction becomes a discounting PGG (dPGG, probability $p$) with diminishing marginal returns to cooperation, or a synergistic PGG (sPGG, probability $1-p$) with increasing marginal returns. Crucially, cooperation elevates $p$ while defection reduces it, establishing a closed feedback loop where strategies alter their future payoff environment.} 
	\label{model}
\end{figure}

\begin{table}[width=.9\textwidth,cols=3,pos=ht]
\centering
\caption{Main parameters used in this work.}
\renewcommand{\arraystretch}{1.5}
\setlength{\tabcolsep}{3pt}
\begin{tabular}{l p{0.68\linewidth}}
\hline
\textbf{Symbol} & \textbf{Interpretation} \\
\hline
$N$                 & Group size \\
$n_\mathrm{C}$      & Number of cooperators in a group \\
$r$                 & Multiplication factor in the PGG \\
$c$                 & Cost of cooperation in the PGG \\
$\omega$            & Nonlinear parameter in the nonlinear PGG \\
$\delta$            & Nonlinear coefficient used for sPGG and dPGG \\
$p$                 & Stochastic probability of playing dPGG \\
$x$                 & Fraction of cooperators in the population \\
$\epsilon$          & Relative speed of environmental feedback \\
$\theta$            & Sensitivity of the environment to cooperation and defection \\
$\delta_s$          & Nonlinear coefficient in sPGG \\
$\delta_d$          & Nonlinear coefficient in dPGG \\
\hline
\end{tabular}
\label{tab:symbols}
\end{table}

In an infinite well-mixed population, all individuals interact with each other with equal probability. Let $x$ $(0<x<1)$ denote the fraction of cooperators in the population, such that $1 - x$ represents the fraction of defectors. For any focal individual, the probability that the remaining $N - 1$ group members include $n_\mathrm{C}$ cooperators is given by
\begin{align}\label{eq_g}
	g(n_{\mathrm{C}},N,x)={\binom{N-1}{n_{\mathrm{C}}}} x^{n_{\mathrm{C}}}(1-x)^{N-1-n_{\mathrm{C}}}.
\end{align}
Accordingly, the average payoffs for cooperators and defectors are
\begin{equation}\label{eq_Pi_well}
\begin{aligned}
\Pi_{\mathrm{C}}(x,p) &= \sum_{n_{\mathrm{C}}=0}^{N-1} g(n_{\mathrm{C}},N,x)\pi_{\mathrm{C}}(n_{\mathrm{C}}+1,p) \nonumber\\
&= \frac{r}{N} \left[ \frac{1 - p}{\delta_s} \left( (1 + \delta_s)(1 + \delta_s x)^{N-1} - 1 \right) + \frac{p}{\delta_d} \left( 1 - (1 - \delta_d)(1 - \delta_d x)^{N-1} \right) \right] - 1, \\
\Pi_{\mathrm{D}}(x,p) &= \sum_{n_{\mathrm{C}}=0}^{N-1} g(n_{\mathrm{C}},N,x) \pi_{\mathrm{D}}(n_{\mathrm{C}},p) \nonumber\\
&= \frac{r}{N} \left[ \frac{1 - p}{\delta_s} \left( (1 + \delta_s x)^{N-1} - 1 \right) \ + \frac{p}{\delta_d} \left( 1 - (1 - \delta_d x)^{N-1} \right) \right].
\end{aligned}
\end{equation}

Furthermore, the temporal evolution of the fraction of cooperators $x$ in the population follows the replicator dynamics, which can be expressed as
\begin{equation}\label{eq_dynamic_x}
	\begin{aligned}
		\dot{x}&=x(1-x)[\Pi_{\mathrm{C}}(x,p)-\Pi_{\mathrm{D}}(x,p)]  \\
               &=x(1-x)\{ \frac{r}{N}\left[(1-p)(1+\delta_s x)^{N-1} + p(1-\delta_d x)^{N-1}\right]-1 \},
	\end{aligned}
\end{equation}
where \( \bar{\Pi}(x, p) = x \Pi_\mathrm{C}(x, p) + (1 - x) \Pi_\mathrm{D}(x, p) \) denotes the average payoff in the population.

Since different values of the parameter $p$ correspond to different game scenarios, we regard $p$ as a representation of the game environment. We assume that the environmental parameter $p$ is influenced by the population state and is regulated by a centralized third party. As the proportion of defectors in the population increases, the probability of encountering an sPGG, given by $1 - p$, increases to promote cooperation. Conversely, as the fraction of cooperators increases, the game environment improves, but due to resource limitations, the probability of a dPGG, denoted by $p$, also increases. Moreover, changes in $p$ directly modify the structure of the game, thereby affecting the composition of cooperators and defectors in the population (see Fig.~\ref{model}).

Here, we consider a linear feedback mechanism: the parameter $p$ increases with the fraction of cooperators and decreases with the fraction of defectors. Accordingly, the dynamical equation governing $p$ is given by
\begin{align}\label{eq_dynamic_p}
    \dot p=\epsilon p(1-p)[\theta x-(1-x)],
\end{align}
where $\epsilon > 0$ denotes the relative speed of environmental feedback. $\theta > 0$ represents the ratio between the enhancement rate due to cooperators and the degradation rate due to defectors, capturing the sensitivity of the feedback to the level of cooperation. The direction of change in $p$ is completely determined by the function $f(x) = \theta x - (1 - x)$. The linear feedback mechanism captures the first-order kinetics of environmental interactions, where the regeneration rate of environmental resources is directly proportional to the abundance of cooperators~\cite{wang2025coevolutionary}. We adopt this minimal model to ensure methodological rigor. Since our primary objective is to isolate the effects of payoff nonlinearity, introducing a complex feedback function with additional parameters would lead to parameter confounding. Thus, adopting the linear feedback ensures that the system's complex co-evolutionary patterns are driven by the intrinsic coupling structure between the game and the environment, rather than by the specific form of the feedback function.

The overall structure of our model is illustrated in Fig.~\ref{model} and can be formally expressed as

\begin{equation}\label{eq_dynamic_well}
    \begin{cases}
        \dot x&=x(1-x)\{ \frac{r}{N}\left[(1-p)(1+\delta_s x)^{N-1} + p(1-\delta_d x)^{N-1}\right]-1 \},\\
        \dot p&=\epsilon p(1-p)[\theta x-(1-x)],
    \end{cases}
\end{equation}

To facilitate a clearer understanding of all the parameters used in our study, we summarize them in Table~\ref{tab:symbols}.

\begin{figure}
    \centering
    \includegraphics[width=0.85\textwidth]{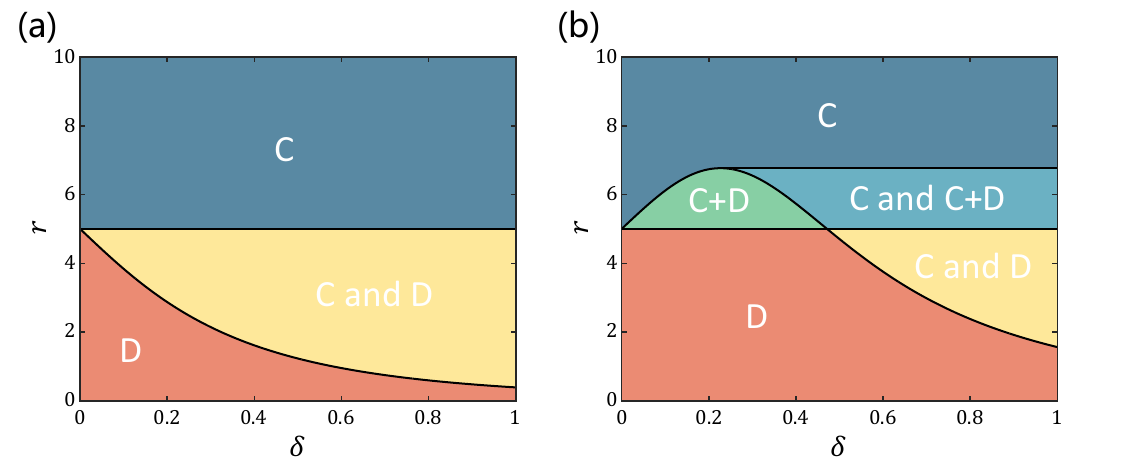}
	\caption{\textbf{The phase diagrams of the $r-\delta$ parameter plane with fixed $p$}. When $r<N$, only two outcomes occur across $\delta$: full defection (D) or bistability between full cooperation and full defection (C/D). When $r>N$, the steady-state type depends on $p$: for small p in panel (a), most of the space converges to full cooperation (C), whereas for large p in panel (b), additional regimes appear, including interior coexistence (C+D) and mixed bistability between C and C+D. Color coding: red denotes D; yellow denotes C/D bistability; blue denotes C; green denotes C+D; light blue denotes C/C+D bistability. Parameters: $N=5$; (a) $p=0.2$; (b) $p=0.8$.} 
	\label{fig:fixed_p}
\end{figure}

\section{Results}\label{sec3}

In this section, we combine theoretical derivations with numerical simulations to systematically analyze the eco-evolutionary dynamics of the nonlinear public goods game. We begin by establishing an evolutionary baseline based on theoretical analysis in a static environment, rigorously delineating the steady-state structures across different parameter regimes. Building on this foundation, we introduce the dynamical feedback mechanism and employ stability analysis alongside numerical computations to reveal how the sensitivity and speed of environmental feedback drive the transitions between stable coexistence and periodic oscillations. Finally, we extend the theoretical framework to the asymmetric case, further exploring how the differential configuration of synergistic and discounting effects reshapes the system's long-term behavior.

\begin{figure}
    \centering
    \includegraphics[width=1\textwidth]{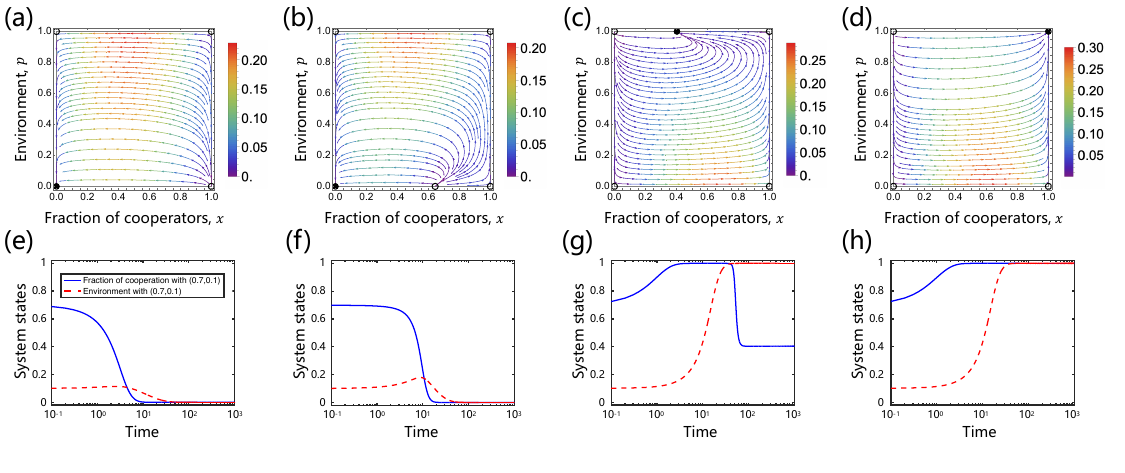}
	\caption{\textbf{Eco-evolutionary dynamics when no interior equilibrium exists}. (Top row) Phase portraits showing vector fields (arrows) and streamlines in the $x-p$ plane, with color scale indicating flow speed magnitude. (Bottom row) Corresponding temporal dynamics from initial condition $(x_0,p_0)=(0.7,0.1)$, with cooperator frequency $x(t)$ (solid blue) and environmental state $p(t)$ (dashed red). Common parameters: $N=5$, $\theta=2$, $\epsilon=0.1$. Subplot-specific parameters: (a,e) $r=1$, $\delta=0.4$; (b,f) $r=2$, $\delta=0.4$; (c,g) $r=7$, $\delta=0.2$; (d,h) $r=9$, $\delta=0.1$.} 
	\label{boundary}
\end{figure}

\subsection{Evolutionary Dynamics in a fixed environment}

We first examine the dynamical behavior of the system under a fixed parameter $ p $. Consistent with previous studies, the system exhibits five qualitatively distinct long-term outcomes, classified by their asymptotic states. The analytical derivations are provided in~\cite{jiang2025nonlinearpublicgoodsgame}, while the classification results are summarized in the $ r $–$ \delta $ phase diagrams shown in Fig.~\ref{fig:fixed_p}.

When $ r < N $, only two types of equilibria are observed. Specifically, if $r < \frac{N}{(1 - p)(1 + \delta)^{N - 1} + p(1 - \delta)^{N - 1}}$, the population evolves towards all-defection. Conversely, if $r > \frac{N}{(1 - p)(1 + \delta)^{N - 1} + p(1 - \delta)^{N - 1}}$, the system exhibits the bistability between all-cooperation and all-defection.

When $r > N$, the number and type of stable outcomes depend on the value of $p$. For $p<0.5 $, the system always converges to all-cooperation. For $p>0.5$, the dynamics becomes more intricate. If  $r<\frac{N}{(1 - p)(1 + \delta)^{N - 1} + p(1 - \delta)^{N - 1}}$,  the population reaches a state of the coexistence of cooperation and defection. If $r > \frac{N}{(1 - p)(1 + \delta)^{N - 1} + p(1 - \delta)^{N - 1}}$, the outcome depends on the sign of the payoff difference function $h(x) = \Pi_C(x, p) - \Pi_D(x, p) = \frac{r}{N}\left[(1 - p)(1 + \delta x)^{N - 1} + p(1 - \delta x)^{N - 1}\right] - 1$. Let $h'(x)=0$, then we have $x^{**}:=\frac{k-1}{\delta(k+1)}$, where $k:=\Big(\tfrac{p}{1-p}\Big)^{\!\frac{1}{N-2}}$. If $ \delta < \delta^* $, or $ \delta > \delta^* $ with $ h(x^{**}) > 0 $, the system converges to all-cooperation, where $\delta^{*}=\frac{k-1}{k+1}$. Otherwise, if $ \delta > \delta^* $ and $ h(x^{**}) < 0 $, it exhibits the bistability between all-cooperation and the coexistence~\cite{jiang2025nonlinearpublicgoodsgame}.

As shown in Fig.~\ref{fig:fixed_p}, a small $ p $ ($ p < 0.5 $) strongly promotes unconditional cooperation (Fig.~\ref{fig:fixed_p}(a)), while a larger $ p $ ($ p > 0.5 $) only partially supports cooperation (Fig.~\ref{fig:fixed_p}(b)).

\subsection{Eco-evolutionary nonlinear PGG with dynamic environments}

By analyzing the system of differential Eq. (\ref{eq_dynamic_well}) when $\delta_s=\delta_d$, we find that the system can have up to seven fixed points. These include four corner equilibrium points: $M_1 = (0,0)$, $M_2 = (1,0)$, $M_3 = (0,1)$, and $M_4 = (1,1)$; two boundary equilibrium points: $M_5 = (x_5, 0) = \left( \frac{e^{\frac{\ln N - \ln r}{N-1}} - 1}{\delta},\ 0 \right)$ and $M_6 = (x_6, 1) = \left( \frac{1 - e^{\frac{\ln N - \ln r}{N-1}}}{\delta},\ 1 \right)$; and one interior equilibrium point: $M_7 = (x_7, n_7) = \left( \frac{1}{\theta + 1},\ \frac{N - r (1 + \delta x_7)^{N-1}}{r \left[ (1 - \delta x_7)^{N-1} - (1 + \delta x_7)^{N-1} \right]} \right)$. In~\cite{jiang2025nonlinearpublicgoodsgame}, we provide a detailed analysis of the existence and stability conditions for all equilibria. In particular, the four corner equilibrium points always exist and, among them, $M_2$ and $M_3$ are always unstable. The boundary equilibrium points $M_5$ and $M_6$ do not coexist, which means that only one can exist for a given set of parameters. In addition, when $M_5$ exists, it is always unstable. In other words, among all possible equilibria, $M_1$, $M_4$, $M_6$, and $M_7$ can potentially be stable under certain parameter regimes. Based on the existence of the interior equilibrium $M_7$, we classify the evolutionary results of the system into two main categories.

We first consider the case in which the interior equilibrium does not exist, that is, when $r < \frac{N}{(1 + \delta x_7)^{N-1}}$ or $r > \frac{N}{(1 - \delta x_7)^{N-1}}$. Under these conditions, the system exhibits three distinct monostable outcomes. When $r < N$, the all-defection equilibrium $M_1$, which corresponds to an environmentally depleted state, becomes the only stable equilibrium, as shown in Fig.~\ref{boundary}(a) and Fig.~\ref{boundary}(e). In particular, when $e^{\frac{\ln N - \ln r}{N-1}} - 1 < \delta < 1$, the boundary point $M_5$ acts as an unstable saddle, and the system still evolves towards $M_1$. This evolutionary trajectory is illustrated in Fig.~\ref{boundary}(b) and Fig.~\ref{boundary}(f). In both scenarios, only the sPGG regime persists in the long term. When $1 - e^{\frac{\ln N - \ln r}{N-1}} < \delta < 1$, the boundary equilibrium $M_6$ becomes the only stable point, representing the coexistence of cooperation and defection in a rich environment. Fig.~\ref{boundary}(c) shows that all initial conditions in the phase plane eventually converge to $M_6$, while all four corner equilibria are unstable. Taking $(0.7, 0.1)$ as an example, Fig.~\ref{boundary}(g) demonstrates that the population first evolves towards the coexistence of cooperators and defectors, followed by environmental enrichment, where all individuals participate in dPGG. Finally, as shown in Fig.~\ref{boundary}(d) and Fig.~\ref{boundary}(h), when $N < r(1 - \delta)^{N-1}$, the system evolves toward the corner point $M_4$, corresponding to full cooperation in a fully enriched environment where only dPGG is present.

\begin{figure}
    \centering
    \includegraphics[width=0.85\textwidth]{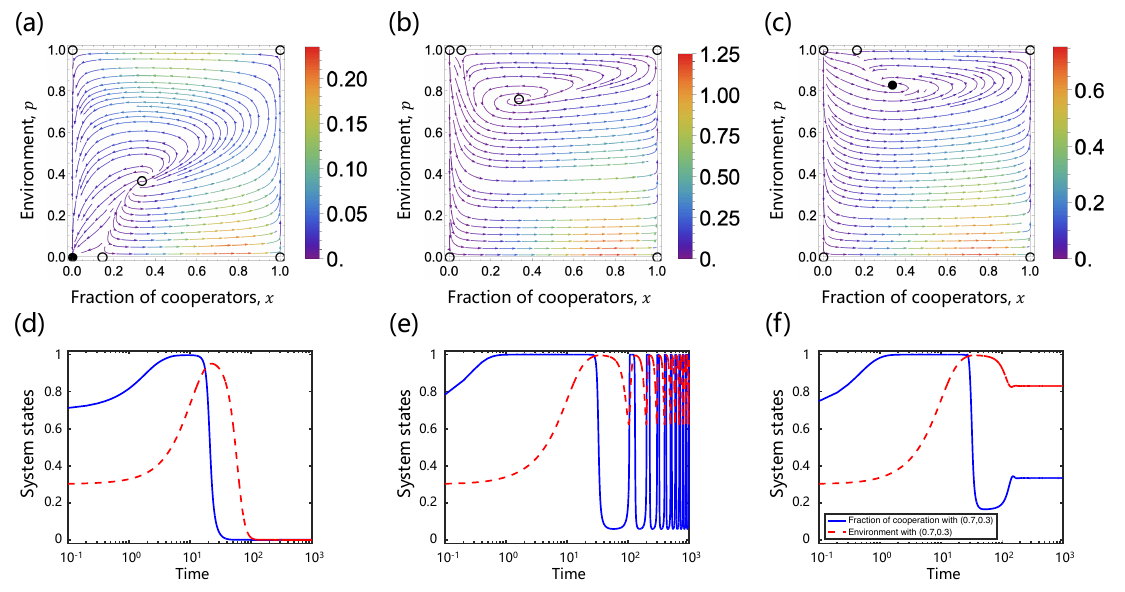}
	\caption{\textbf{Eco-evolutionary dynamics with interior equilibria}. Similar to Fig.~\ref{boundary}, top row shows phase portraits in the x–p plane, and bottom row shows corresponding temporal dynamics from initial condition $(x_0,p_0)=(0.7,0.3)$. The panels illustrate the three outcome types: (a,d) Corner attraction: convergence to $M_1$ (full defection) with an unstable interior equilibrium; (b,e) persistent oscillations: a stable interior limit cycle; (c,f) interior stability: stabilization at $M_7$ with intermediate levels of cooperation and environment. Common parameters: $N=5$, $\theta=2$, $\epsilon=0.1$. Subplot-specific parameters: (a,d) $r=4$, $\delta=0.4$; (b,e) $r=6$, $\delta=0.8$; (c,f) $r=7$, $\delta=0.5$.} 
	\label{interior}
\end{figure}

We now turn to the three types of stable outcomes that arise when the interior equilibrium exists, that is, for $ \frac{N}{(1 + \delta x_7)^{N-1}} < r < \frac{N}{(1 - \delta x_7)^{N-1}} $. First, when $ \frac{N}{(1 + \delta x_7)^{N-1}} < r < N $, the only stable equilibrium is $M_1$. However, the phase plane also contains several unstable equilibria: the vertex points $M_2$, $M_3$, and $M_4$, the boundary point $M_5$, and the interior point $M_7$, as illustrated in Fig.~\ref{interior}(a) and Fig.~\ref{interior}(d). Next, when $N < r < \frac{1}{2} \left[ \frac{N}{(1 - \delta x_7)^{N-2}} + \frac{N}{(1 + \delta x_7)^{N-2}} \right]$, Fig.~\ref{interior}(b) shows that none of the equilibria is stable, including $M_1$, $M_2$, $M_3$, $M_4$, $M_6$, and $M_7$. Instead, a limit cycle emerges in the interior of the phase plane. In other words, as shown in Fig.~\ref{interior}(e), the cooperation level and environmental state of the population exhibit sustained periodic oscillations. Finally, when $ \frac{1}{2} \left[ \frac{N}{(1 - \delta x_7)^{N-2}} + \frac{N}{(1 + \delta x_7)^{N-2}} \right] < r < \frac{N}{(1 - \delta x_7)^{N-1}} $, the interior equilibrium $M_7$ becomes the only stable point, accompanied by an unstable boundary point $M_6$ and all four unstable vertex points. Fig.~\ref{interior}(c) and Fig.~\ref{interior}(f) reveal that, under this regime, the system stabilizes at intermediate values of both cooperation level and environmental state, with neither approaching 0 nor 1.

\begin{figure}
    \centering
    \includegraphics[width=0.6\textwidth]{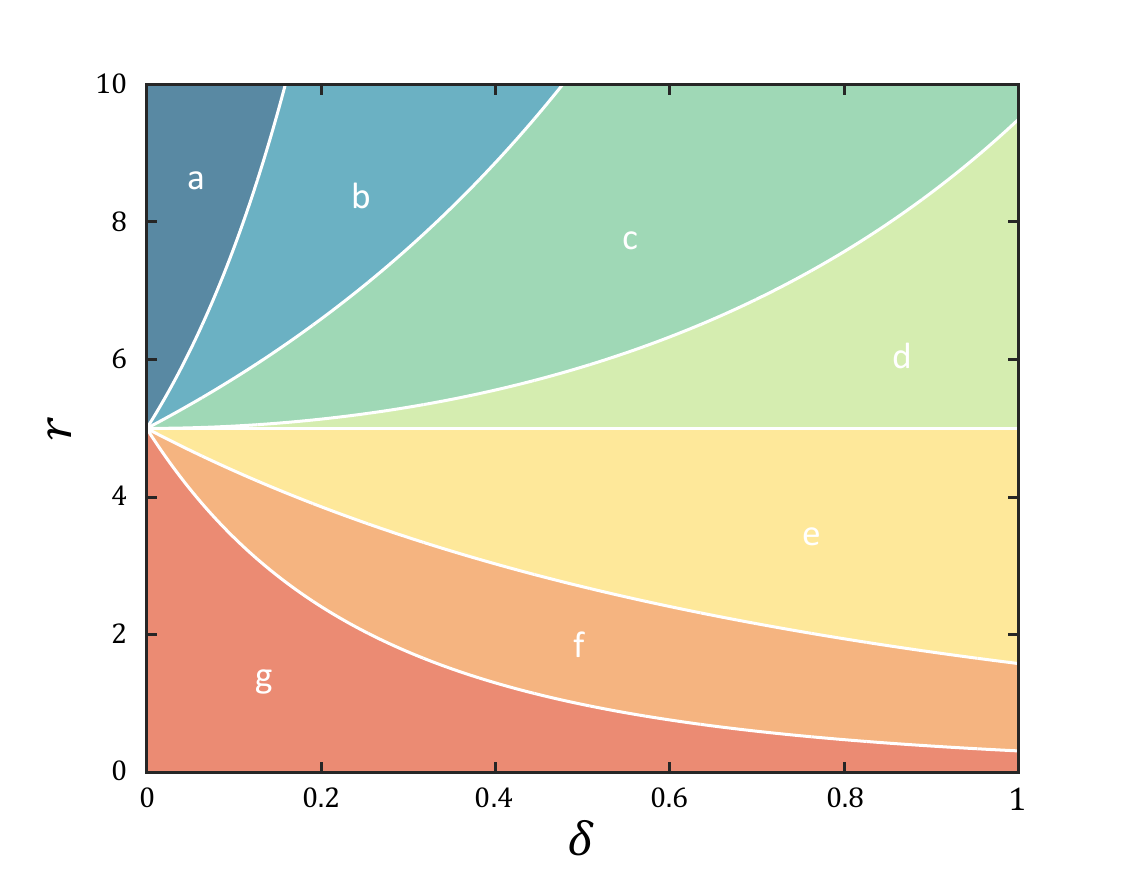}
	\caption{\textbf{The phase diagrams of the $r-\delta$ parameter plane with dynamic environments}. The parameter space is partitioned by analytic boundaries (white curves), with the horizontal line at $r=N$ providing a fundamental demarcation. For $r<N$ (regions e, f and g), the system evolves toward $M_1$, representing full defection in a depleted environment. For $r>N$, four distinct regimes emerge: (a) full cooperation $M_4$ in an enriched environment; (b) boundary coexistence $M_6$; (c) interior fixed point $M_7$ with intermediate cooperation and environment ($x_7=1/(1+\theta)$); and (d) interior limit cycle. Parameters: $N=5$, $\theta=2$.} 
	\label{phase}
\end{figure}

To summarize all the possible steady states discussed above, we use the $r$–$\delta$ phase diagram (Fig.~\ref{phase}). First, in the lower half-plane where $r < N$, regions e, f, and g in Fig.~\ref{phase} all feature $M_1$ as the only stable equilibrium point. This implies that the population evolves toward full defection under a purely sPGG regime. In particular, all three regions contain three unstable corner equilibria. Furthermore, region f also includes an unstable boundary equilibrium $M_5$, while region e contains both the unstable $M_5$ and an unstable interior point $M_7$. In the upper half-plane where $r > N$, four regions (a, b, c, and d) are characterized by different dynamical outcomes. In region a, the system converges to full cooperation in a purely dPGG environment, where $M_4$ is stable and the remaining three vertex points are unstable. Region b corresponds to the case in which the boundary equilibrium $M_6$ is stable, representing the stable coexistence of cooperation and defection under dPGG, with a long-term cooperation level of $x_6$. In region c, the interior equilibrium $M_7$ is the only stable point. Here, the population stabilizes at the cooperation level of $x_7 = 1/(1+\theta)$, with $p_7$ being the long-term probability of playing dPGG and $1 - p_7$ that of sPGG. Finally, in region d, the cooperation level and environmental state undergo persistent oscillations between 0 and 1, indicating the emergence of a limit cycle in the system.

\begin{figure}
    \centering
    \includegraphics[width=0.85\textwidth]{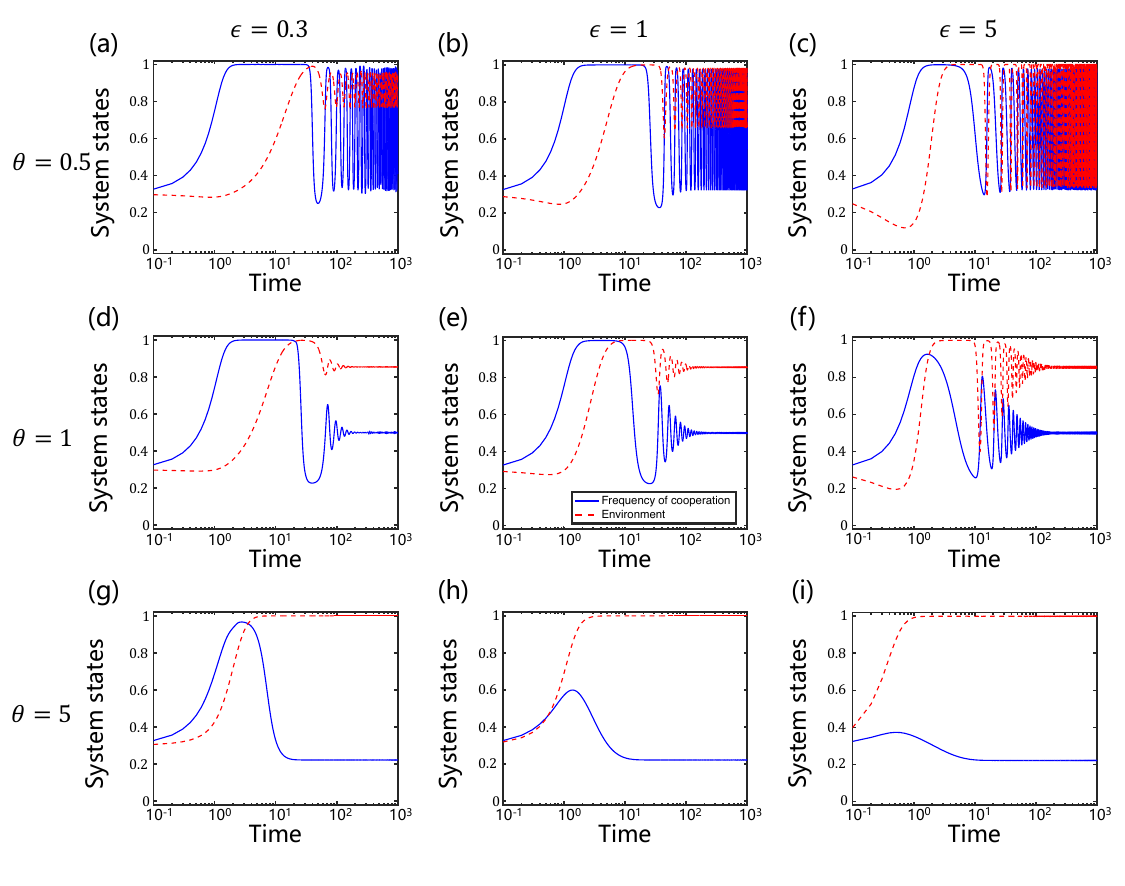}
	\caption{\textbf{Influence of environment sensitivity to cooperation and defection $\theta$ and the relative speed of environmental feedback $\epsilon$}. Each panel plots the cooperator fraction $x(t)$ (blue, solid) and the environmental state $p(t)$ (red, dashed) from the same initial condition $(x_0,p_0)=(0.3,0.3)$. Rows fix $\theta\in\{0.5,1,5\}$ and columns fix $\epsilon\in\{0.3,1,5\}$. Panels (a)-(c) show that the trajectories settle into an interior limit cycle , with larger $\epsilon$ producing greater oscillation amplitude in $p$ and shorter transients. The system converges to the interior fixed point $M_7$ in panels (d)-(f), and the dynamics approach the boundary equilibrium $M_6$ in panels (g)-(i), with faster convergence and stronger early excursions as $\epsilon$ increases. The fixed parameters are $N=5$, $r=8$ and $\delta=0.5$.}
	\label{theta_epsilon}
\end{figure}

Then, we investigate the effects of the environmental feedback rate $\epsilon$ and the environmental sensitivity to cooperation, indicated by $\theta$. From the previous analysis, we observed that richer dynamical behaviors tend to emerge when $r > N$. Therefore, we fix $r = 8 > N = 5$ for this section. To capture the influence of feedback speed, we consider $\epsilon = 0.3$, $1$, and $5$. To reflect different levels of environmental sensitivity or preference towards cooperation, we examine $\theta = 0.5$, $1$, and $5$. Each row in Fig.~\ref{theta_epsilon} corresponds to a fixed value of $\theta$, while each column corresponds to a fixed value of $\epsilon$. As shown in Fig.~\ref{theta_epsilon} , the system exhibits three distinct types of long-term behavior under this parameter configuration: an interior limit cycle (region d in Fig.~\ref{phase}), a stable interior equilibrium $M_7$ (region c in Fig.~\ref{phase}), and a stable boundary equilibrium $M_6$ (region b in Fig.~\ref{phase}). When the environmental preference for cooperation is weak (that is, $\theta$ is small), both the cooperation level and the environmental state exhibit sustained oscillations between 0 and 1 (see Fig.~\ref{theta_epsilon}(a)–(c)).  When $\theta$ is moderate ($\theta = 1$) or large ($\theta = 5$), the system stabilizes at the interior equilibrium $M_7$ or the boundary equilibrium $M_6$, respectively, as illustrated in Fig.~\ref{theta_epsilon}(d)–(f) and Fig.~\ref{theta_epsilon}(g)–(i). This indicates that $\theta$ plays a determining role in the type of steady state attained. 

Furthermore, $\epsilon$ mainly affects the amplitude and convergence speed of the system's transient dynamics. For example, Fig.~\ref{theta_epsilon}(a)–(c) show that larger values of $\epsilon$ lead to greater oscillation amplitudes in the environmental variable $p$, while the fluctuation range of the cooperation level $x$ remains relatively unchanged. In addition, $\epsilon$ also influences the intensity of transient oscillations before reaching equilibrium. As shown in Fig.~\ref{theta_epsilon}(f) compared to Fig.~\ref{theta_epsilon}(d), the system exhibits more pronounced transient fluctuations when $\epsilon$ is higher. Lastly, larger values of $\epsilon$ also accelerate convergence to the final steady state, as observed in Fig.~\ref{theta_epsilon}(g)–(i). The physical intuition behind this phenomenon stems from the timescale separation between environmental feedback and evolutionary adaptation. A larger $\epsilon$ implies that the environmental response to population behavior is significantly faster than the strategic adaptation. When minor fluctuations in cooperation occur a high $\epsilon$ drives the environmental state to adjust rapidly. This fast environmental response propels the system onto wider orbits in the phase space thereby manifesting as large-amplitude oscillations. Conversely a smaller $\epsilon$ imparts an inertial character to the environment causing its variation to lag behind strategic adjustments and thus constraining the system within a smaller oscillation range.

\subsection{Asymmetric nonlinear PGG with environmental feedback}

In the previous section, we discussed a symmetric nonlinear PGG payoff structure, where both dPGG and sPGG employed the same nonlinear coefficient $\delta$. Here, we focus on asymmetry in dPGG and sPGG.

\begin{figure}
    \centering
    \includegraphics[width=0.8\textwidth]{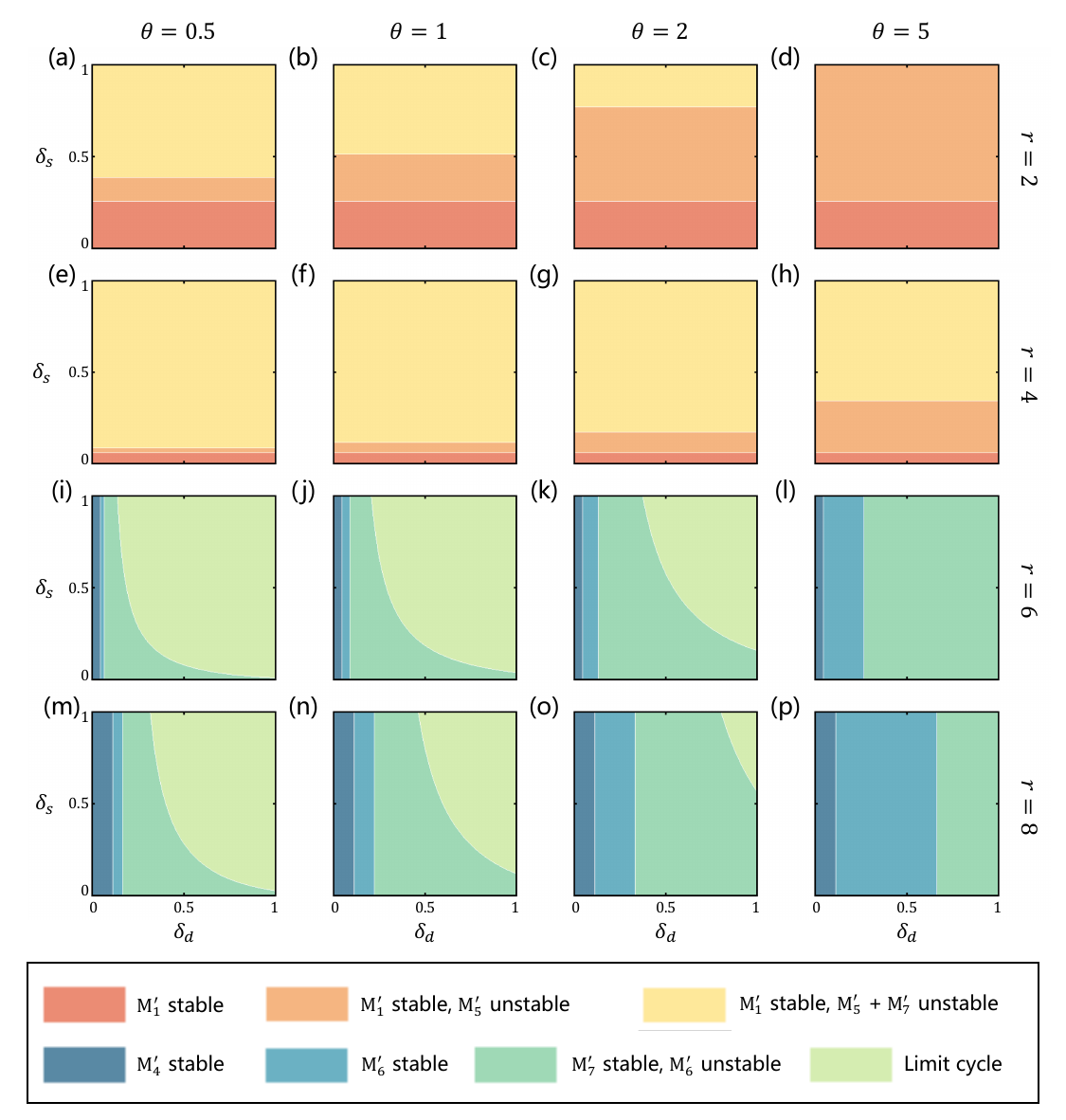}
    \caption{\textbf{Outcome maps on the $\delta_d-\delta_s$ plane under asymmetric nonlinearities.} Columns vary sensitivity of environment $\theta = \{0.5, 1, 2, 5\}$ and rows vary multiplication factor $r = \{2, 4, 6, 8\}$. Axes show discounting ($\delta_d$, horizontal) and synergy ($\delta_s$, vertical) nonlinearities. Colors follow the legend below. For $r < N$ (top two rows), the defection attractor $M_1'$ dominates, with unstable boundary or interior equilibria. For $r > N$ (bottom two rows), outcomes shift systematically with $\delta_d$: small $\delta_d$ favors full cooperation ($M_4'$); moderate $\delta_d$ stabilizes boundary coexistence ($M_6'$); the system then transitions within interior steady states—from a stable interior fixed point $M_7'$ to an oscillatory attractor (limit cycle) as $\delta_d$ increases. Particularly, increasing either $r$ or $\theta$ compresses the oscillatory regime. Parameter: $N=5$.}
    \label{delta_phase}
\end{figure}

We solve Eq. ~\ref{eq_dynamic_well} to determine the equilibrium points of the system. As in the previous case, up to seven equilibrium points may exist. These include four corner equilibria: $M_1' = (0, 0)$, $M_2' = (1, 0)$, $M_3' = (0, 1)$, and $M_4' = (1, 1)$; two boundary equilibria: $M_5' = (x_5', 0) = \left( \frac{e^{\frac{\ln N - \ln r}{N-1}} - 1}{\delta_s},\ 0 \right)$ and $M_6' = (x_6', 1) = \left( \frac{1 - e^{\frac{\ln N - \ln r}{N-1}}}{\delta_d},\ 1 \right)$; and one interior equilibrium: $M_7' = (x_7', n_7') = \left( \frac{1}{\theta + 1},\ \frac{N - r(1 + \delta_s x_7')^{N-1}}{r \left[ (1 - \delta_d x_7')^{N-1} - (1 + \delta_s x_7')^{N-1} \right]} \right)$. In~\cite{jiang2025nonlinearpublicgoodsgame}, we provide the mathematical derivation of the existence and stability conditions for these equilibria. Similarly to the previous analysis, $M_2'$ and $M_3'$ are always unstable, and $M_5'$, if it exists, is also unstable. The existence and stability of the remaining equilibria depend on the specific values of the parameters. In Fig.~\ref{delta_phase}, we explore how the steady-state structure depends on parameters $\delta_d$, $\delta_s$, $\theta$ and $r$. The horizontal axis represents $\delta_d$ and the vertical axis represents $\delta_s$. Each row of the figure corresponds to a fixed value of $r$, while each column corresponds to a fixed value of $\theta$. Different regions in the phase diagram are color-coded to indicate different types of steady states, and the color scheme corresponds one-to-one with that used in Fig.~\ref{phase}.

First, the relationship between $r$ and $N$ (with $N = 5$) determines the dominant strategy underlying the steady-state outcome. Specifically, when $r < N$, the system favors defection and is mainly influenced by $\delta_s$ (Fig.~\ref{delta_phase}(a)–(h)). In contrast, when $r > N$, the system tends to favor cooperation and is shaped mainly by $\delta_d$ (Fig.~\ref{delta_phase}(i)–(p)).

For the case of $r < N$, as the multiplication factor $r$ increases, the region in the $\delta_d$–$\delta_s$ phase plane where the boundary equilibrium $M_5'$ exists expands (the orange and yellow areas in Fig.~\ref{delta_phase}(a) and (e)). Additionally, the existence region of the interior equilibrium $M_7'$ is positively correlated with $r$ (the yellow regions in Fig.~\ref{delta_phase}(d) and (h)), but negatively correlated with the sensitivity of the level of cooperation and defection $\theta$ (the yellow regions in Fig.~\ref{delta_phase}(a)–(d)). In the case of $r > N$, higher values of $r$ lead to expansion of the stable regions of $M_4'$ and $M_6'$ in the $\delta_d$–$\delta_s$ plane. The stability region of $M_6'$ also increases with increasing $\theta$. The stable region of $M_7'$, together with the region where the limit cycle exists, constitutes the interior dynamical regime. This interior region shrinks with increasing values of $\theta$ and $r$. In particular, the existence of a limit cycle is nonlinearly regulated by both $\delta_d$ and $\delta_s$, and it appears when $N < r < \frac{N}{\delta_s + \delta_d} \left[ \frac{\delta_s}{(1 - \delta_d x_7^{'})^{N - 2}} + \frac{\delta_d}{(1 + \delta_s x_7^{'})^{N - 2}} \right].$
This indicates that larger values of $\delta_d$ and $\delta_s$ tend to facilitate the emergence of limit cycles. Furthermore, smaller values of $r$ and $\theta$ result in a wider region for the existence of limit cycles (the light green regions in Fig.~\ref{delta_phase}(i)–(k) and (m)–(o)), whereas larger $r$ and $\theta$ can directly suppress or eliminate the oscillatory behavior (Fig.~\ref{delta_phase}(l) and (p)).

In sum, asymmetry assigns distinct jobs. On the one hand, synergy acts as an amplifier when conditions are defection-prone, boosting the payoff of small cooperative clusters. On the other hand, discounting serves as a tuner once cooperation is viable, deciding whether the system settles near coexistence or keeps oscillating.

\section{Conclusion and Discussion}\label{sec4}
The environmental state plays a critical role in shaping the evolution of complex social systems, influencing both individual strategies and long-term collective dynamics~\cite{wang2020eco}. Previous research mainly adopted static environments in nonlinear games or simple linear feedback mechanisms to investigate the evolution of cooperation in linear games~\cite{ito2024complete}. Although these studies have uncovered fundamental patterns of cooperative behavior, they have not fully captured the intricate feedback mechanisms between environmental states and cooperative dynamics in nonlinear games that occur in real-world scenarios. To address this gap, we introduce a dynamic environmental feedback mechanism into a nonlinear PGG, systematically examining the coevolution of environmental states and cooperative strategies. Our model integrates the stochastic combination of synergistic (sPGG) and discounting (dPGG) nonlinear interactions, revealing diverse dynamic behaviors such as full defection, full cooperation, stable coexistence, and periodic limit cycles driven by the joint effects of environmental feedback, nonlinearity, and stochasticity. These findings significantly extend the theoretical framework of traditional PGG models and offer novel insights into cooperative dynamics within real-world social, biological, and economic systems.

Furthermore, under dynamic environmental feedback conditions, we explored the eco-evolutionary dynamics in a nonlinear PGG, uncovering more complex and diverse dynamic phenomena compared to previous studies that considered static or fixed environmental randomness. Explicitly incorporating dynamic interactions between environmental states and cooperative strategies, our model identified up to seven equilibrium points, encompassing various outcomes such as complete defection, complete cooperation, coexistence of cooperation and defection, and periodic limit cycles. In particular, we provided detailed analyses of how the environmental feedback rate $\epsilon$ and the sensitivity of the environment to cooperation $\theta$ influence the evolution. Smaller values of $\theta$ were found to induce periodic fluctuations between cooperation levels and environmental states, while larger values of $\theta$ facilitated stable states of cooperation or coexistence. Furthermore, the feedback rate $\epsilon$ significantly affected the speed of reaching equilibrium and the amplitude of the transitional oscillations. These results not only highlight the crucial role of dynamic environmental feedback in cooperative evolution but also systematically quantify the influences of environmental sensitivity and feedback speed, offering critical theoretical and practical insights for understanding complex interactions between environments and cooperative behaviors.

Moreover, this study further introduced asymmetric nonlinear parameters $\delta_s$ and $\delta_d$ to more precisely distinguish between sPGG and dPGG interactions, a distinction rarely explored in previous studies. Through a comprehensive analysis of equilibrium existence and stability under asymmetric conditions, we elucidated how different steady-state structures change with variations in the multiplication factor $r$, nonlinear coefficients $\delta_s$ and $\delta_d$, and environmental sensitivity $\theta$. Our results indicated that when $r<N$, the system tends to stabilize in states dominated by defection, primarily influenced by $\delta_s$; conversely, when $r>N$, cooperation-dominated states become more prevalent, strongly affected by $\delta_d$. In addition, we systematically revealed that larger values of $\delta_s$ and $\delta_d$ tend to induce periodic oscillations, while larger values of $r$ and $\theta$ can suppress or even eliminate these oscillations. This asymmetric analysis not only better aligns with real-world cooperative dynamics but also enriches our theoretical understanding of how nonlinearity and stochasticity jointly affect cooperation.

Stochasticity is an essential factor to consider when studying complex systems~\cite{gao2024learning, borner2024saddle, zeng2023temporal, kleshnina2023effect}. Recent research conceptualized stochasticity as uncertainty between nonlinear synergistic and discounting public goods games~\cite{zhu2024evolutionary}. Our work further extends this model by expanding the randomness from static to scenarios of continuous dynamics. The main contribution of our research lies in systematically integrating randomness, nonlinearity, and dynamic environmental feedback, thereby uncovering complex dynamics in cooperative evolution. Our findings overcome the theoretical limitations inherent in traditional PGG models and provide practical implications for cooperative management in real-world scenarios~\cite{wang2024paradigm}. Particularly, in the management of charitable resource flows~\cite{kastelic2024cooperation}, the public willingness to donate forms a co-evolutionary relationship with the transparency or efficiency of resource allocation. Policymakers can avoid drastic fluctuations in donation enthusiasm by regulating the sensitivity of information feedback thereby maintaining the long-term stability of the system in a high-synergy state. Analogously in managing large-scale health crises public compliance with intervention measures and disease prevalence exhibit similar coupled dynamical characteristics~\cite{weitz2020awareness}. Policymakers must carefully regulate the responsiveness of interventions to prevent the system from falling into recurrent epidemic waves thereby maintaining a stable state of containment. Broadly speaking the framework constructed in this study provides a general theoretical tool for modelling environmental feedback loops which is crucial for understanding the coupled dynamics of socio-ecological systems~\cite{shuvo2024investigating}.

However, due to the inherent complexity of real-world systems, our study has certain limitations and potential avenues for future research. It is important to note that the analysis in this paper is based on the assumption of an infinite well-mixed population and adopts a linear form to characterize environmental feedback. Relaxing these settings to consider sophisticated nonlinear environmental feedback mechanisms~\cite{wang2020steering}, structured populations~\cite{liu2024evolutionary, wang2024evolutionary, sun2025feedback}, and group reputation~\cite{schmid2021unified, murase2023indirect} is expected to reveal richer evolutionary steady states. Additionally, the cross-scale evolution of cooperation, integrating micro-level individual interactions and macro-level population dynamics, represents a promising direction for uncovering universal cooperation patterns~\cite{lamarins2022importance, kawakatsu2024stereotypes}. Notably, conducting human-computer interaction experiments to validate and deepen the practical applicability and effectiveness of theoretical models is also an important future research direction~\cite{makovi2023trust, zhu2025capturing}.

\section*{CRediT authorship contribution statement}
\textbf{Yishen Jiang}: Methodology, Formal analysis, Conceptualization, Investigation, Writing – original draft, Writing – review \& editing; \textbf{Xin Wang}: Methodology, Formal analysis, Investigation, Writing – review \& editing, Funding acquisition; \textbf{Wenqiang Zhu}: Methodology, Formal analysis, Investigation, Writing – review \& editing; \textbf{Ming Wei}: Investigation, Writing – review \& editing; \textbf{Longzhao Liu}: Methodology, Formal analysis, Investigation, Writing – review \& editing, Funding acquisition; \textbf{Shaoting Tang}: Investigation, Writing – review \& editing, Funding acquisition; \textbf{Hongwei Zheng}: Investigation, Writing – review \& editing, Funding acquisition.

\section*{Data availability}
No data was used for the research described in the article.

\section*{Declaration of Competing Interest}
The authors declare that they have no known competing financial interests or personal relationships that could have appeared to influence the work reported in this paper.

\section*{Acknowledgements}
This work is supported by National Science and Technology Major Project (2022ZD0116800), Program of National Natural Science Foundation of China (12425114, 62141605, 12201026, 12301305, 62441617, 12501702), the Fundamental Research Funds for the Central Universities, Beijing Natural Science Foundation (Z230001), National Cyber Security-National Science and Technology Major Project (2025ZD1503700), the Opening Project of the State Key Laboratory of General Artificial Intelligence (Project No. SKLAGI2025OP16), and Bejing Advanced Innovation Center for Future Blockchain and Privacy Computing.

\appendix
\section*{Appendix A. Evolutionary dynamics of stochastic nonlinear PGG in a static environment}
Fix the environment at a constant $p\in[0,1]$. The population state $x\in[0,1]$ follows the one–dimensional replicator equation
\[
\dot x \;=\; x(1-x)\,h(x),
\]
where $h(x)\;=\;\frac{r}{N}\Big[(1-p)(1+\delta x)^{N-1}+p(1-\delta x)^{N-1}\Big]-1.$ Then $h(0)=\frac{r}{N}-1$, and $h(1)=\frac{r}{N}\,A(\delta)-1$, where $A(\delta)=(1-p)(1+\delta)^{N-1}+p(1-\delta)^{N-1}$.
 
Differentiating gives
\[
h'(x)=\frac{r}{N}(N-1)\delta\Big[(1-p)(1+\delta x)^{N-2}-p(1-\delta x)^{N-2}\Big].
\]
Let $h'(x)=0$, then we have $x^{**}:=\frac{k-1}{\delta(k+1)}$, where $k:=\Big(\tfrac{p}{1-p}\Big)^{\!\frac{1}{N-2}}$. If $x<x^{**}$, $h'(x)<0$ and $h(x)$ decreases monotonically; $x>x^{**}$, $h'(x)>0$ and $h(x)$ increases monotonically.

Next we discuss the effect of $p$ and $\delta$ on the monotonicity of $h(x)$:
\begin{itemize}
\item If $p\le \tfrac12$, we have $x^{**}\le 0$ and $h(x)$ is increasing monotonically on $[0,1]$.
\item If $p>\tfrac12$ and $\delta>\delta^{*}=\frac{k-1}{k+1}$, then $x^{**}\in(0,1)$ and $h(x)$ decreases on $(0,x^{**})$ and increases on $(x^{**},1)$.
\item If $p>\tfrac12$ and $\delta<\delta^{*}$, then $x^{**}>1$ and $h(x)$ decreases monotonically on $(0,1)$.
\end{itemize}

Now we can analyze stable states of the system:

\noindent (1) \textbf{$r<N$} ($h(0)<0$)
\begin{itemize}
\item If $r< \frac{N}{A(\delta)}$ then $h(1)<0$ and the only attractor is all-defection ($x=0$).
\item If $\frac{N}{A(\delta)}<r<N$ then $h(0)<0<h(1)$. The dynamics are bistable between all-defection ($x=0$) and all-cooperation ($x=1$), separated by an unstable interior point.
\end{itemize}

\noindent (2) \textbf{$r>N$} ($h(0)>0$)
\begin{itemize}
\item If $p\le\tfrac12$, then $x^{**}\le 0$, hence all-cooperation ($x=1$) is globally stable.
\item If $p>\tfrac12$ and $x^{**} > 0$:

(a) If $r<\frac{N}{A(\delta)}$ then $h(1)<0$. There is an internal stable point where cooperation and defection can coexist.

(b) If $r>\frac{N}{A(\delta)}$ then $h(1)>0$
\begin{itemize}
\item If $\delta<\delta^{*}$, or $\delta>\delta^{*}$ and $h(x^{**})>0$, all-cooperation is the only stable state.
\item If $\delta>\delta^{*}$ and $h(x^{**})<0$, the dynamics are bistable between all-cooperation and coexistence with an unstable internal fixed point. 
\end{itemize}
\end{itemize}

\section*{Appendix B. Eco-evolutionary dynamics with environmental feedback}
We study the symmetric case with environmental feedback where the nonlinear strength is identical in sPGG and dPGG, denoted by $\delta\in(0,1)$. Let $x\in[0,1]$ be the fraction of cooperators and $p\in[0,1]$ the environmental state, interpreted as the probability of playing dPGG. The coevolutionary dynamics are
\begin{equation}
 \begin{cases}
        \dot x&=x(1-x)H(x,p)\\
        \dot p&=\epsilon p(1-p)[\theta x-(1-x)]
    \end{cases},
\end{equation}
where
\[
H(x,p) = \tfrac{r}{N}\bigl[(1-p)(1+\delta x)^{N-1}+p(1-\delta x)^{N-1}\bigr]-1,
\]
and the rectangle $\mathcal{D}=[0,1]\times[0,1]$ is positively invariant.

The Jacobian $J = \begin{bmatrix} a_{11} & a_{12} \\ a_{21} & a_{22} \end{bmatrix}$ has entries
\begin{align*}
a_{11}&=\frac{\partial \dot x}{\partial x} \;=\; (1-2x)\,H(x,p)\;+\;x(1-x)\,H_x(x,p),\\
a_{12}&=\frac{\partial \dot x}{\partial p} \;=\; x(1-x)\,H_p(x,p),\\
a_{21}&=\frac{\partial \dot p}{\partial x} \;=\; \epsilon\,p(1-p)\,(\theta+1),\\
a_{22}&=\frac{\partial \dot p}{\partial p} \;=\; \epsilon\,(1-2p)\,\big[(\theta+1)x-1\big],
\end{align*}
with 
{\small
\begin{align*}
H_x(x,p) &= \frac{r}{N}(N-1)\delta\bigl[(1-p)(1+\delta x)^{N-2}-p(1-\delta x)^{N-2}\bigr],\\
H_p(x,p) &= \frac{r}{N}\bigl[(1-\delta x)^{N-1}-(1+\delta x)^{N-1}\bigr]<0.
\end{align*}
}

By solving $\dot x=0$ and $\dot p=0$, we can obtain up to seven equilibrium points of the system.

(1) Corner equilibrium $M_1=(0,0)$
\[J(0,0)=\begin{bmatrix} \frac{r}{N}-1 & 0 \\ 0 & -\epsilon \end{bmatrix}.\]
Eigenvalues are $\lambda_1=r/N-1$ and $\lambda_2=-\epsilon<0$. Hence $M_1$ is stable iff $r<N$ (when $\lambda_1<0$).

(2) Corner equilibrium $M_2=(1,0)$
\[J(1,0)=\begin{bmatrix} 1-\frac{r(1+\delta)^{N-1}}{N} & 0 \\ 0 & \epsilon\theta \end{bmatrix}.\]
Because there is a positive eigenvalue $\lambda_2=\epsilon\theta>0$, $M_2$ is always unstable.

(3) Corner equilibrium $M_3=(0,1)$
\[J(0,1)=\begin{bmatrix} \frac{r}{N}-1 & 0 \\ 0 & \epsilon \end{bmatrix}.\]
Because there is a positive eigenvalue $\lambda_2=\epsilon>0$, $M_3$ is always unstable.

(4) Corner equilibrium $M_4=(1,1)$
\[J(1,1)=\begin{bmatrix} 1-\frac{r(1-\delta)^{N-1}}{N} & 0 \\ 0 & -\epsilon\theta \end{bmatrix}.\]
Eigenvalues are $\lambda_1=1-\frac{r(1-\delta)^{N-1}}{N}$ and $\lambda_2=-\epsilon\theta<0$. Hence $M_4$ is stable if $r>\frac{N}{(1-\delta)^{N-1}}$ (when $\lambda_1<0$).

(5) Boundary equilibrium $M_5=(x_5=\frac{e^{\frac{\ln N - \ln r}{N-1}} - 1}{\delta},0)$

The existence condition of $M_5$ is $0<\frac{e^{\frac{\ln N - \ln r}{N-1}} - 1}{\delta}<1$, which is equal to$\frac{N}{(1+\delta)^{N-1}}<r<N$.
\[J(x_5,0)=\begin{bmatrix} a_{11}(x_5,0) & a_{12}(x_5,0) \\ 0 & \epsilon((\theta+1)x_5-1) \end{bmatrix}.\]
Here, $\lambda_1=x_5(1-x_5)H_x(x_5,0)=\frac{x_5(1-x_5)\delta (N-1)}{1+\delta x_5}>0$, so $M_5$ is never stable if it exists.

(6) Boundary equilibrium $M_6=(x_6=\frac{1-e^{\frac{\ln N - \ln r}{N-1}} }{\delta},1)$

The existence condition of $M_6$ is $0<\frac{1-e^{\frac{\ln N - \ln r}{N-1}} }{\delta}<1$, which is equal to $N<r<\frac{N}{(1-\delta)^{N-1}}$. It means that $M_5$ and $M_6$ can not coexist.
\[J(x_6,1)=\begin{bmatrix} a_{11}(x_6,1) & a_{12}(x_6,1) \\ 0 & -\epsilon((\theta+1)x_6-1) \end{bmatrix}.\]
$\lambda_1=x_6(1-x_6)H_x(x_6,0)=-\frac{x_6(1-x_6)\delta (N-1)}{1-\delta x_6}<0$ and $\lambda_2=-\epsilon((\theta+1)x_6-1)$. If $\lambda_2<0$, we can get $x_6 > \frac{1}{1+\theta}$. Therefore $M_6$ is stable if $\frac{N}{(1-\frac{\delta}{\theta+1})^{N-1}}<r<\frac{N}{(1-\delta)^{N-1}}$).

(7) Interior equilibrium $M_7=(x_7=\frac{1}{1+\theta},p_7)$

Here, $p_7$ is uniquely determined by $H(x_7,p_7)=0$:
\[p_7 \;=\; \frac{\frac{N}{r}-(1+\delta x_7)^{N-1}}{(1-\delta x_7)^{N-1}-(1+\delta x_7)^{N-1}}.\]
The existence condition of $M_7$ is $0<p_7<1$, that is, $\frac{N}{(1+\delta x_7)^{N-1}} \;<\; r \;<\; \frac{N}{(1-\delta x_7)^{N-1}}$.
\[J(x_7,p_7)=\begin{bmatrix} a_{11}(x_7,p_7) & a_{12}(x_7,p_7) \\ a_{21}(x_7,p_7) & 0 \end{bmatrix}.\]
The fixed point $M_7$ is stable if and only if the following two conditions hold:
\[
\begin{cases}
    \det (J(M_7))>0\\
    \operatorname{tr}(J(M_7))<0
\end{cases}.
\]
First, $\det (J(M_7))=-a_{12}(x_7,p_7)a_{21}(x_7,p_7)=-\,x_7(1-x_7)\,H_p(x_7,p_7)\,\epsilon\,p_7(1-p_7)(\theta+1) \;$. Because $H_p(x_7,p_7)<0$ , $\det (J(M_7))>0$ always holds. Second, $\operatorname{tr}(J(M_7))=a_{11}(x_7,p_7)=\; x_7(1-x_7)\,H_x(x_7,p_7)$. If we want $\operatorname{tr}(J(M_7))<0$, we need to make $H_x(x_7,p_7)<0$, that is, $r>r^{*}=\frac{1}{2}\left[\frac{N}{(1-\delta x_7)^{N-2}}+\frac{N}{(1+\delta x_7)^{N-2}}\right]$, and we can prove that $r^{*}<\frac{N}{(1-\delta x_7)^{N-1}}$ always hold. As a consequence, $M_7$ is stable iff $r^{*}<r<\frac{N}{(1-\delta x_7)^{N-1}}$.

However, by summarizing the conclusions above, we can find that there is no stable fixed point when $N<r<r^{*}$, although four corner fixed points, boundary fixed point $M_6$ and interior fixed point $M_7$ exist. The domain $\mathcal{D}=[0,1]\times[0,1]$ is a positively invariant bounded closed set, because on the boundary $x=0$, $x=1$, $p=0$, and $p=1$ the components of the vector field vanish (i.e., $\dot x=0$ or $\dot p=0$), hence trajectories cannot leave $\mathcal{D}$. All equilibria are unstable, and for any interior initial condition (that is, $x\in(0,1)$ and $p\in(0,1)$) the $\omega$-limit set of the trajectory cannot contain any equilibrium (since equilibria are unstable and repelling). By the Poincaré--Bendixson theorem, if a trajectory is contained in a compact set that has no equilibria, then its $\omega$-limit set is a periodic orbit. Here, although $\mathcal{D}$ contains boundary equilibria, the $\omega$-limit set of interior trajectories actually lies in the interior (because boundary equilibria are unstable and are not approached by trajectories). Therefore, at least one limit cycle exists.

\section*{Appendix C. Eco-evolutionary dynamics of asymmetric nonlinearities}
Now, we study the asymmetric case where the nonlinear strength differs between sPGG and dPGG. The coevolutionary dynamics are
\begin{equation*}
 \begin{cases}
        \dot x&=x(1-x)G(x,p)\\
        \dot p&=\epsilon p(1-p)[\theta x-(1-x)]
    \end{cases},
\end{equation*}
Here, for the asymmetric case, we denote the payoff difference function by $G(x,p)$ to distinguish it from the symmetric function $H(x,p)$ introduced in Appendix B:
\begin{equation*}
G(x,p)\;=\;\frac{r}{N}\Big[(1-p)\,(1+\delta_s x)^{N-1}\;+\;p\,(1-\delta_d x)^{N-1}\Big]\,-\,1.
\end{equation*}

The Jacobian is $J = \begin{bmatrix} a_{11} & a_{12} \\ a_{21} & a_{22} \end{bmatrix}$, where
\begin{align*}
a_{11}&=\frac{\partial \dot x}{\partial x} \;=\; (1-2x)\,G(x,p)\;+\;x(1-x)\,G_x(x,p),\\
a_{12}&=\frac{\partial \dot x}{\partial p} \;=\; x(1-x)\,G_p(x,p),\\
a_{21}&=\frac{\partial \dot p}{\partial x} \;=\; \epsilon\,p(1-p)\,(\theta+1),\\
a_{22}&=\frac{\partial \dot p}{\partial p} \;=\; \epsilon\,(1-2p)\,\big[(\theta+1)x-1\big],
\end{align*}
with 
{\small
\begin{align*}
G_x(x,p) &= \frac{r}{N}(N-1)\Bigl[(1-p)\delta_s(1+\delta_s x)^{N-2}-p\delta_d(1-\delta_d x)^{N-2}\Bigr],\\
G_p(x,p) &= \frac{r}{N}\Bigl[(1-\delta_d x)^{N-1}-(1+\delta_s x)^{N-1}\Bigr].
\end{align*}
}

Similarly, we can also obtain up to seven equilibrium points of the system by solving $\dot x=0$ and $\dot p=0$.

(1) Corner equilibrium $M_1'=(0,0)$
\[
J(0,0)=\begin{bmatrix} \tfrac{r}{N}-1 & 0 \\ 0 & -\epsilon \end{bmatrix}.
\]
Eigenvalues: $\lambda_1=r/N-1,\;\lambda_2=-\epsilon<0$. Hence $M_1'$ is stable iff $r<N$.

(2) Corner equilibrium $M_2'=(1,0)$
\[
J(1,0)=\begin{bmatrix} 1-\tfrac{r(1+\delta_s)^{N-1}}{N} & 0 \\ 0 & \epsilon\theta \end{bmatrix}.
\]
Since $\lambda_2=\epsilon\theta>0$, $M_2'$ is always unstable.

(3) Corner equilibrium $M_3'=(0,1)$
\[
J(0,1)=\begin{bmatrix} \tfrac{r}{N}-1 & 0 \\ 0 & \epsilon \end{bmatrix}.
\]
Since $\lambda_2=\epsilon>0$, $M_3'$ is always unstable.

(4) Corner equilibrium $M_4'=(1,1)$
\[
J(1,1)=\begin{bmatrix} 1-\tfrac{r(1-\delta_d)^{N-1}}{N} & 0 \\ 0 & -\epsilon\theta \end{bmatrix}.
\]
Eigenvalues: $\lambda_1=1-\tfrac{r(1-\delta_d)^{N-1}}{N}$, $\lambda_2=-\epsilon\theta<0$. Hence $M_4'$ is stable if $r>\tfrac{N}{(1-\delta_d)^{N-1}}$.

(5) Boundary equilibrium $M_5'=\bigl(x_5'=\tfrac{e^{\frac{\ln N - \ln r}{N-1}} - 1}{\delta_s},0\bigr)$
\[J(x_5',0)=\begin{bmatrix} a_{11}(x_5',0) & a_{12}(x_5',0) \\ 0 & \epsilon((\theta+1)x_5'-1) \end{bmatrix}.\]
It exists iff $\tfrac{N}{(1+\delta_s)^{N-1}}<r<N$, and it is always unstable since $\lambda_1=a_{11}(x_5')>0$.

(6) Boundary equilibrium $M_6'=\bigl(x_6'=\tfrac{1-e^{\frac{\ln N - \ln r}{N-1}}}{\delta_d},1\bigr)$
\[J(x_6',1)=\begin{bmatrix} a_{11}(x_6',1) & a_{12}(x_6',1) \\ 0 & -\epsilon((\theta+1)x_6'-1) \end{bmatrix}.\]
$M_6'$ exists if $N<r<\tfrac{N}{(1-\delta_d)^{N-1}}$.  
It is stable if $x_6'>\tfrac{1}{1+\theta}$, which gives the condition $\tfrac{N}{(1-\frac{\delta_d}{\theta+1})^{N-1}}<r<\tfrac{N}{(1-\delta_d)^{N-1}}$.

(7) Interior equilibrium $M_7'=(x_7=\tfrac{1}{1+\theta},p_7')$

Here $p_7'$ is determined by $G(x_7,p_7')=0$:
\[
p_7'=\frac{\tfrac{N}{r}-(1+\delta_s x_7)^{N-1}}{(1-\delta_d x_7)^{N-1}-(1+\delta_s x_7)^{N-1}}.
\]
The interior fixed point exists when $0<p_7'<1$, that is $\frac{N}{(1+\delta_s x_7)^{N-1}} \;<\; r \;<\; \frac{N}{(1-\delta_d x_7)^{N-1}}$.  
\[J(x_7',p_7')=\begin{bmatrix} a_{11}(x_7',p_7') & a_{12}(x_7',p_7') \\ a_{21}(x_7',p_7') & 0 \end{bmatrix}.\]
Since $G_p(x_7',p_7')<0$, we can get $\det(J(M_7'))=-a_{12}(x_7',p_7')a_{21}(x_7',p_7')>0$. The trace is $\operatorname{tr}(J(M_7'))=a_{11}(x_7',p_7')=x_7'(1-x_7')G_x(x_7',p_7')$. So $M_7'$ is stable iff $G_x(x_7',p_7')<0$, which holds for $r^{*}(\delta_s,\delta_d)<r<\frac{N}{(1-\delta_d x_7)^{N-1}}$, with $r^{*}(\delta_s,\delta_d) = \frac{N}{\delta_s + \delta_d} \left[ \frac{\delta_s}{(1 - \delta_d x_7^{'})^{N - 2}} + \frac{\delta_d}{(1 + \delta_s x_7^{'})^{N - 2}} \right]$. Similarly, when $N<r<r^{*}(\delta_s,\delta_d)$, a limit cycle appears.

Importantly, the asymmetric framework in Appendix C is a strict generalization of the symmetric case. When setting $\delta_s=\delta_d=\delta$, all theoretical results reduce exactly to those obtained in Appendix B.

\printcredits

\bibliographystyle{elsarticle-num}
\bibliography{ref}
\end{document}